\documentstyle[sprocl]{article}

\input{psfig}
\def\mycite{\,\cite}

\def\be{\begin{equation}}
\def\ee{\end{equation}}
\def\bea{\begin{eqnarray}}
\def\eea{\end{eqnarray}}

\def\ma{m_{\rm a}}
\def\fa{f_{\rm a}}

\def\dalemb#1#2{{\vbox{\hrule height .#2pt
\hbox{\vrule width.#2pt height#1pt \kern#1pt\vrule width.#2pt}
\hrule height.#2pt}}}
\def\square{\mathord{\dalemb{5.9}{6}\hbox{\hskip1pt}}}
\def\tdot{\kern -8.5pt{}^{{}^{\hbox{...}}}}
\def\dotprime{\kern -8.0pt{}^{{}^{\hbox{.}~\prime}}}

\def\lapp{\hbox{$ {     \lower.40ex\hbox{$<$}
                   \atop \raise.20ex\hbox{$\sim$}
                   }     $}  }
\def\gapp{\hbox{$ {     \lower.40ex\hbox{$>$}
                   \atop \raise.20ex\hbox{$\sim$}
                   }     $}  }

\def\marbul{\strut\vadjust{\kern-2pt$\bullet$}}

\def\rr{\rangle}
\mathchardef\less="321C
\def\ll{\langle}
\def\smallskip{\vskip 10pt}
\def\microeV{\mu\hbox{\rm eV}}
\def\meV{\hbox{\rm meV}}
\def\GeV{\hbox{\rm GeV}}

\begin{document}

\title{RECENT PERSPECTIVES ON AXION COSMOLOGY}

\author{R.A. Battye}

\address{Theoretical Physics Group, Blackett Laboratory, Imperial College \\
Prince Consort Road, London, SW7 2BZ, U.K.}

\author{E.P.S. Shellard}

\address{Department of Applied Mathematics and Theoretical Physics \\
University of Cambridge \\ Silver Street, Cambridge, CB3 9EW, U.K.}

\maketitle\abstracts{We review current cosmological constraints on 
the axion.
We describe the basic mechanisms by which axions are created in the early
universe, focussing on the standard thermal scenario where the 
dominant process is through axion 
radiation by a string network.  A dark matter axion
in this case would have a mass $\ma \sim
100\,\microeV$, with specified large uncertainties.  This cosmological 
lower bound leaves a viable window for the axion below
the astrophysical upper limit, $\ma \,\lapp\, 1\hbox{--}10\,\meV$.  We also 
discuss alternative axion cosmologies which allow a much wider, but 
indefinite, mass range.
}
  
\section{Introduction}

Ever since the early 
1980's, the axion has consistently remained one of 
the most popular dark matter candidates.   
Unlike many other more exotic particles,
the axion's existence depends only on a minimal extension of the standard
model\mycite{PQ,WilWei} which also solves one of its key difficulties---the strong CP
problem of QCD.  
In the standard thermal scenario for the early universe, cosmic axions are
created by a variety of mechanisms ranging from `quiescent' production
of zero momentum axions
when their mass `switches on' at $t\approx10^{-6}$ seconds\mycite{PWW} 
through to 
`topological' production by the violent radiative decay of a network of 
axion strings formed at about
$t=10^{-25}$ seconds\mycite{Davb}.  Originally, the role of axion strings 
was overlooked and calculations of the present axion density just from 
`quiescent' mechanisms suggested a very light mass near 
$m_a \approx 5\,\microeV$, that is, if the axion were to solve the dark matter 
problem\mycite{PWW}.  However, it was soon
recognized that axion strings provide the dominant cosmological 
contribution\mycite{Davb,DSa},
and subsequent calculations\mycite{BaSb} 
showed that a heavier mass range was appropriate $m_a\sim 100\,\microeV$,
though with additional uncertainties given the difficulty of these 
nonlinear calculations.  Astrophysical constraints on the axion mass 
suggest that it must be less than about $1\hbox{--}10\,\meV$, so the 
allowable axion parameter window is viable, though relatively narrow.
  
The present large-scale axion search experiment\mycite{axsearch} is
looking near the original mass range, $m_a \sim 1\hbox{--}10\,\microeV$,
for a variety of historical and technological reasons.  Such an 
axion is by no means excluded in some alternative scenarios
which we shall also discuss.  However, this paper will
focus on the `standard thermal scenario' which provides strong
motivation for an expanded quest in search of a slightly heavier axion---an 
experiment that may become technically feasible in the 
not-too-distant future.

\subsection{The origin of the axion}

The standard model of particle physics, based around the Weinberg-Salam 
model for electroweak interactions and QCD for strong interactions, has 
one significant flaw---the strong CP problem: non-perturbative effects 
of instantons add an extra term to the perturbative Lagrangian
and the coefficient of this term, denoted $\bar\theta$, governs the level
 of CP violation in QCD. The absence of any such violation in all observed 
strong interactions imposes a constraint on $\bar\theta$, the most stringent 
being $\bar\theta<10^{-10}$ due to the absence of a neutron electric dipole 
moment\mycite{neutrondipole}. Since the value of $\bar\theta$ is effectively 
arbitrary, one is left with a severe fine tuning problem.

The elegant solution of Peccei and Quinn\mycite{PQ} is to allow $\theta$ to become a 
dynamical field which relaxes toward the CP conserving value $\bar\theta=0$ 
\footnote{In fact, $\bar\theta\ne 0$, since there are some CP violating weak 
interactions, and it has been shown that $\bar\theta>10^{-14}$ in the standard
model\mycite{MW}, although this is highly model dependent.}, by the 
spontaneous breaking of a $U(1)$-symmetry. The resulting particle, known as the 
axion\mycite{WilWei}, has couplings to ordinary matter which are 
proportional to $\fa^{-1}$ and acquires a mass $\ma$ also proportional 
to $\fa^{-1}$ at the QCD phase transition. Initially, it was supposed 
that the Peccei-Quinn scale was close to the electroweak phase transition
$\fa\sim T_{\rm EW}$, but an exhaustive search of accelerator 
data ruled this possibility out, implying $\fa\gapp 10^{7}{\rm GeV}$. 
However, the 
ensuing disappointment was short-lived because there is no phenomenological 
reason why the Peccei-Quinn scale $\fa$ could not be much higher, even up to 
grand-unification scales. Thus, the `invisible' axion was born,
an extremely light particle with almost
undetectably weak couplings of order $\fa^{-1}$.

\subsection{Astrophysical constraints on the axion}

Accelerator limits on the new `invisible' axion were soon superseded by 
astrophysical calculations of cooling rates of large stars, such as 
red giants.  Axions, being weakly coupled, escape from the whole volume
of a star and, in certain parameter ranges, this can exceed the usual 
heat loss mechanisms through convection and other surface effects.  
The oft-quoted red giant constraint is $\fa\gapp 10^{9}{\rm GeV}$ (or 
$\ma \lapp 6\,\meV$)\mycite{redgiant}.  
More stringent constraints on $\fa$, however, were thought to be 
imposed by neutrino emissions from supernova 1987a, providing the 
limit, $\fa\gapp 10^{10}
{\rm GeV}$, which corresponds to a mass bound $\ma \lapp 0.6\,\meV$\mycite{SN}.  
This `conventional' wisdom has been questioned 
by Raffelt {\it et al.}\ who performed a detailed re-examination 
of these astrophysical calculations. 
They suggest that one cannot impose a meaningful constraint
from SN1987a, since there is insufficient data, and, moreover, that
the constraint from red giants is  weaker than  
previously thought, $\fa\gapp 5\times 10^{8}{\rm GeV}$ 
(or $\ma \lapp 10\,\meV$)\mycite{raf}.     
More recent work endeavours to answer 
some of these criticisms and reasserts a somewhat weaker bound from SN1987a,
$\fa \gapp 10^9$GeV (or $\ma \lapp 6\,\meV$), though this 
can be an order of magnitude tighter 
depending on the axion model parameters\mycite{KJSSTE}.  This topic lies
outside the scope of the present discussion, so more general reviews
should be consulted\mycite{axreviews}.

\subsection{Cosmological axions: `quiescent' vs.\ `topological' production}

The weak coupling to ordinary matter and the substantial redshifting 
between the Peccei-Quinn and QCD phase transitions make the 
axion an ideal cold dark matter candidate.
The earliest estimates of the cosmological axion density\mycite{AS,PWW} 
assumed `quiescent' production, that is, axions were created 
through coherent oscillations about the minimum of the Peccei-Quinn
potential when the axion mass `switched on' at $T\sim\Lambda_{\rm QCD}$. 
A recent estimate of the relative contribution of these zero 
momentum axions is
\be
\Omega_{{\rm a},{\rm h}}  \approx 0.9 h^{-2}\Delta
\bigg{(}{\fa\over 10^{12}{\rm GeV}}\bigg{)}^{1.18}  \bar\theta^2_i \,,
\label{homcont}
\ee
where $\Delta\sim 1$ accounts for the uncertainties in the QCD phase 
transition (discussed later)\mycite{Turn}. Assuming an ill-defined
`average' value of the axion field $\langle \bar\theta^2_i\rangle = \pi^2/3$, 
implies a constraint $\fa\lapp 10^{12}{\rm GeV}\,,\ma\gapp 5\mu {\rm eV}$
when compared to the closure density of a flat universe $\Omega=1$.
As we have emphasised already, 
this estimate ignores much stronger topological effects arising at the 
Peccei-Quinn phase transition, that is, the inevitable creation of
an axion string network. 

If we suppose that we live in a universe that underwent an inflationary 
phase in the early universe, then there are two basic scenarios for 
cosmological axion production.  These depend on the relationship 
between the inflationary reheat temperature $T_{\rm reh}$ and the
string-forming phase transition temperature $T_{\rm PQ} \sim \fa$:

\begin{enumerate}

	\item If the reheat temperature $T_{\rm reh}$ is
sufficient to restore the Peccei-Quinn symmetry $T_{\rm reh}>\fa$, then
a network of global or axion strings will form by the Kibble mechanism.
This is the usual thermal scenario commonly assumed in most of the axion 
literature.
Here, we {\it can} predict the mass of the axion if it constitutes the main dark
matter component of the universe.

	\item Alternatively, if $T_{\rm reh}<\fa$, then any strings which may 
have been formed before the inflationary epoch would be diluted 
by the subsequent rapid exponential expansion with which it is associated.
In this case, we {\it cannot}, in principle,  make a definite prediction
about the axion mass on cosmological grounds.

\end{enumerate}

\noindent It is important to elaborate this `predictive' distinction because, 
in simple inflationary scenarios with $T_{\rm reh} < \fa$,  
the estimate (\ref{homcont}) will still 
apply and a mass prediction appears to be meaningful. 
However, $\theta_{\rm i}$ is homogeneous over an entire inflationary 
domain, certainly exceeding the present horizon, and it is set to a fixed, but 
arbitrary, value. By (\ref{homcont}), this {\it unknown} value 
of $\theta_{\rm i}$ in our domain
very precisely defines the value of $\ma$ required 
for the axion to be the dark matter---or vice versa.  Either way,
we have some sort of tuning re-emerging, which is not 
particularly consistent with the original motivation for the 
axion\footnote{Linde suggests that in the 
infinite `manifold' of a  chaotic inflationary model,  
all values of $\theta_{\rm i}$ will be realized, so this fine-tuning 
can be interpreted merely as a manifestation of the 
anthropic principle\mycite{Lindeb}; the axion-to-baryon ratio 
is set by $\theta_{\rm i}$
and humankind can only tolerate a narrow band of ratios! 
However, many find this form of `explanation' decidedly unappealing.}.
Nevertheless,  in section 4 we discuss this scenario, 
which gives a wide range of 
possible axion masses, along with 
other `non-standard' scenarios.  We also note that topological defects 
may have formed during inflation\mycite{Lyth3} or that they may have been 
created at 
the end of inflation during a phase transition\mycite{Linde}, 
effectively removing this new fine-tuning problem by returning us to the 
first case (i). For this reason, we describe the axion string picture as the 
`standard thermal scenario'; it is essentially the original axion cosmology 
understood more completely.

If  a string network does form, then its decay into axions will provide the dominant contribution
to the overall axion density\mycite{Davb,DSa,BSd}, since the topological 
effects involved are much
stronger than the coherent zero-momentum contributions of
(\ref{homcont}). In this scenario the axion field
$\theta$ is assigned different values at every point in space 
by the topological requirement
that $\theta$ changes by $2\pi$ around each string. This removes the 
possibility of any kind of fine-tuning of $\theta_{\rm i}$, 
since the distribution of strings 
is dictated by the dynamics of the phase transition. However, the spectrum of the
radiation from these strings has been the subject of much
debate\mycite{Dava,Davb,HarS,HagS,DSa} because it dictates the magnitude of this
contribution. Indeed, it appears that the  over-production of axions by strings
could almost close the allowed window of values for $\fa$, given the lower
bound provided by astrophysical effects. 

In recent years, we have undertaken a thorough investigation of  the
string radiation spectrum, using both analytic and numerical
techniques\mycite{BSa,BSc,BSd,Ba}. The findings were in 
broad agreement with the initial work of 
Davis {\it et al.}\mycite{Dava,Davb,DSa}, but not with 
the work of Sikivie {\it et al}\mycite{HarS,HagS} which results in a weaker
constraint. Taken at face value the earlier results of Davis {\it et al.}\ 
would eliminate the axion as a cold dark matter candidate 
in the standard scenario. However, although the
underlying physics of the earlier work was correct,  the model for 
string evolution on which it was based was too simplistic.  Subsequent work
has provided a more sophisticated picture of axion production by a string
network which does leave an window open for the axion.

\subsection{Overview}

In this article, we review the cosmological constraints on the axion, 
in particular, adding greater detail to an earlier letter on the 
axion string bound.  We introduce a simple model for the 
evolution of a network of  
cosmic axion strings based on a `one-scale' model. Using this model we
calculate expressions for the contibution to the axion density from 
string loops and the long string network, exhibiting explicit parameter
dependencies.  By comparing to the closure density, we estimate the 
constraint on the Peccei-Quinn symmetry breaking scale $\fa$ and the axion 
mass $\ma$. In the standard scenario, we show that this contribution 
completely dominates the axion
density from `quiescent' zero-momentum production\mycite{PWW}.

We note that there has been considerable controversy associated 
with the nature of
axion string radiation and how this affects the resulting cosmological 
constraint.
For this reason we provide an appendix discussing this topic in some detail.
We present a tight mathematical argument, based on the low energy 
Kalb-Ramond action for axion strings, which demonstrates that radiation 
goes primarily into the lowest frequency modes available. 
This understanding is shown to be 
in quantitative agreement with high resolution numerical simulations of 
the full U(1) field equations even at moderately high energies. This
establishes beyond reasonable doubt the efficacy of the Kalb-Ramond 
action for describing axion string dynamics, thus providing a firm basis 
for the cosmological calculations.

Finally, we discuss `non-standard scenarios' including some of the broad
array of inflationary variants 
and we also consider the possibility of axion dilution  
by entropy production due to the out-of-equilibrium decay of 
massive particles.  Our conclusions summarize the best current estimates 
of the cosmological constraint on the axion.

\section{Axion string network evolution}

\noindent The evolution of axion strings is qualitatively very similar to the
evolution of local strings due to their dynamical correspondence---as
demonstrated in the appendix. The
additional long-range field, due to the coupling to the axion, 
acts primarily to renormalize the string
tension and energy density, 
\be
\mu \approx 2\pi \fa^2 \ln
(t/\delta)\,, \label{muren}
\ee
where the string core width is $\delta \sim \fa^{-1}$ and we
assume the typical curvature radius of the strings at a time $t$ is $R\sim t$.
Quantitatively, on small scales $\ell \less t$, global strings are
affected by enhanced  radiation backreaction; typically in a cosmological
context axion radiation will be three orders of magnitude stronger than the weak
gravitational radiation produced by local strings.   This difference will alter
small-scale features such as string wiggliness and loop creation sizes, but not
the more robust large-scale network properties. We comment further on
the nature of string radiation in the next section.

All the pertinent events
take place in the radiation era, so for definiteness we consider a flat ($\Omega
=1$) FRW model with  
\be
a\propto t^{1/2}\,,\quad\rho = {3\over 32\pi Gt^2}\,,\quad
n \approx 0.12 {\cal N}\, T^3\,, \quad t\approx 0.3{\cal
N}^{-1/2}{m_{\rm pl}\over T^2}\,,
\ee
where $\rho$ is the energy density, $n$ is the 
particle number density, 
and ${\cal N}(T)$ is the effective number of massless degrees of freedom at the
temperature $T$.

\subsection{String network formation and the damped epoch}

\noindent The string-forming phase transition creates a tangled network
permeating throughout the universe.  The largest fraction (over 80\%) 
of string makes up a random walk of long
or `infinite' strings, and a scale-invariant distribution of closed
loops makes up the remainder.  After formation, these 
strings experience a significant damping force due to
the relatively high radiation background density.
In the case of local strings, the dominant interaction is through
Aharonov-Bohm type scattering. This frictional force will dominate the 
dynamics of the strings, until the Hubble damping force becomes larger. The temperature
and time corresponding of this transition are given by
\be
T_{*}\sim (G\mu)^{1/2}\eta\,,\quad t_{*}\sim (G\mu)^{-1}t_{\rm f}\,,
\ee
where $G= m_{\rm pl}^{-2}$ is Newton's constant, $\mu$ is the string energy 
per unit length, $\eta$ is the symmetry breaking scale and $t_{\rm f}$ is the 
time of formation.

There are no gauge fields present for global
strings and therefore Aharonov-Bohm type scattering no longer pertains. 
In this case, one finds that Everett scattering is the
dominant process, with a scattering cross-section\mycite{martin}
\be
{d\sigma\over d\theta}={\pi\over 4q[\log(\delta q)]^2}\,,
\ee
where $q\sim T$ is the momentum of the incident particle and
$\delta\sim\fa^{-1}$ is the width of the string.

As for local strings, the frictional force per unit string length can be estimated
as 
\be
{\bf F}\sim n\sigma_{\rm t} v_{T}\Delta{\bf
p}\,,
\ee
where $\sigma_t$ is the transport cross-section given by 
\be
{\sigma_{\rm t}}={\pi^2\over
2q[\log(q\delta)]^2}\sim T^{-1}[\log(T\delta)]^{-2}\,,
\ee
$v_T\sim 1$ is the thermal velocity of the particles and  $\Delta{\bf p}\sim -T{\bf v}$ is
the average momentum transfer per collision where $T$ is the temperature.

By comparison to the Hubble damping force we find that
the temperature at which frictional damping becomes negligible is given by
\be
{{T_*\over[\log(T_*/\fa)]^2}} \sim G\mu m_{\rm pl}\,.
\ee
This implies that
strings oscillate relativistically and begin to radiate axions from the time, 
\be
t_*\sim 10^{-20}\bigg{(}{\fa\over 10^{12}{\rm GeV}}\bigg{)}^{-4}\,{\rm sec}\,.
\ee

\subsection{The scaling regime for string network evolution}

\noindent After the damped epoch, the strings
are expected to evolve towards a scale-invariant regime by the formation of loops
and subsequent emission of radiation. This scaling regime is likely to be achieved
irrespective of the radiative
mechanism, although the important parameters describing the network will be somewhat different. 

The overall density of strings splits neatly into
two distinct parts, the long strings with length $\ell > t$ and small closed
loops with $\ell < t$,
\be
\rho=\rho_{\infty}+\rho_{L}\,.
\ee
The exact length scale $l_c,$ at which
the distinction between long strings and loops is made is unimportant since large
loops are rare, but it is normally assumed to be comparable to the horizon. 
However, this
distinction is important since loops will be radiated  away quickly, whereas long
strings will remain until domain walls form close to the QCD phase transition.

If we ignore radiative effects, then the dynamics of the strings can be 
described by the Nambu action. The equations of motion for a string in an expanding 
background are 
\be
\ddot{\bf X}-{1\over\epsilon}\bigg{(}{{\bf
X}^{\prime}\over\epsilon}\bigg{)}^{\prime}=-{2\dot a\over a}(1-\dot{\bf
X}^2)\dot{\bf X}\,,\quad\dot\epsilon =-{2\dot a\over
a}\epsilon\dot{\bf X}^2\,,\label{stringeom}
\ee
using the temporal transverse gauge, where the string 
coordinates are $X^{\mu}=(t,{\bf X}$), $\dot{\bf X}\cdot{\bf
X}^{\prime}=0$ and $\epsilon^2={\bf X}^{\prime 2}/(1 - \dot {\bf X}^2)$. 
In this gauge, the string energy is given by
\be
E=\mu_0\int d\sigma\epsilon\,,
\ee
where $\sigma$ is a spacelike coordinate along the string.
Differentiating this expression with respect to time and substituting in the expression for $\dot\epsilon$ from (\ref{stringeom}), one can obtain 
\be
\dot E = -2\ll v^2\rr E\,,\quad \dot\rho = -2(1+\ll v^2\rr)\rho\,,
\ee
where $\ll v^2\rr$ is the average string velocity defined by
\be
\ll v^2\rr=\int d\sigma\epsilon \dot{\bf X}^2\bigg{/}\int d\sigma\epsilon\,,
\ee
with $E=\rho V$ and $V\propto a^3$. In this model, $\ll v^2\rr$ is an unknown
constant to be determined for long strings, and $\ll v^2\rr\approx 1/2$
for string loops. Therefore,
excluding radiative effects and the formation of loops,
the density of long strings and loops evolve independently according to
\be
\dot\rho_{\infty}=-{2\dot a\over a}\big{(}1+\ll v^2\rr\big{)}\rho_{\infty}\,,
\qquad\dot\rho_{L}=-{3\dot a\over a}\rho_{L}\,.
\label{noloop}
\ee

The equations (\ref{noloop}) do not take into account the loss of energy to loops. 
The rate of energy loss from the long string network into loops can be described in terms of a
scale invariant production function $f(\ell/L)$ where $\ell$ is the loop size and
$L$ is the characteristic length of the long string network. This 
function is defined so that $\mu f(\ell /L)d\ell /L$ gives the energy loss into loops of
size $\ell$ to $\ell+d\ell$ per unit time per correlation volume $L^3$. The equations for the 
long string density and the loop density are
\be
\dot\rho_{\infty}=-{2\dot a\over a}\big{(}1+\langle v^2\rangle\big{)}\rho_{\infty}
-{c\rho_{\infty}\over L}\,,\qquad
\dot\rho_{L}=-{3\dot a\over a}\rho_{L} + {g\mu\over L^4}f(\ell/L)\,,
\ee
where $g$ is a Lorentz factor and $c$ is measure of loop production rate given by
\be
c\rho_{\infty}={\mu\over L^3}\int_{0}^{\ell_c}d\ell
\,f(\ell/L)\,.
\ee 

In order to investigate these equations, one can 
substitute $\rho_{\infty}=\mu\zeta/t^2$ and $L=\zeta^{-1/2}t$ into the equation 
for $\rho_{\infty}$. If $\zeta$ is constant, then we have that the density of long strings
will scale like radiation or matter in their respective eras, and the evolution 
is self-similar that is, the large scale features of the string network will look the same
at all times, except for a universal scaling proportional to the change in horizon size.
Performing this substitution, one obtains
\be
{\dot\zeta\over\zeta}={1\over t}\bigg{(}2-2\beta\big{(}1+\ll v^2\rr\big{)}
-c\zeta^{1/2}\bigg{)}\,,
\ee
where  the scale factor is given by $a(t)\propto t^{\beta}$. This equation has a stable
fixed point solution for $\zeta$, corresponding to the scaling regime.
Setting the right hand side of this equation 
to zero allows one to derive a relation between the loop production rate, the
long string density and the average velocity of the long strings.

Using $a(t)\propto t^{1/2}$ in the radiation era,
one can deduce that the loop production rate required to maintain scaling is 
$c=\zeta^{-1/2}\big{(}1-\ll v^2\rr\big{)}\,$.
One can 
also solve the equation for the density of loops to give
\be
\rho_{L}={g\mu\zeta^{5/4}\over(\ell t)^{3/2}}\int^{\infty}_{\zeta
^{1/2}\ell/t}dx\sqrt{x}
f(x)\,,
\ee
which can be approximated at late times by $\rho_{L}=\mu\nu (\ell t)^{-3/2}$ 
where 
\be
\nu=g\zeta^{5/4}\int_0^{\infty}dx\sqrt{x}f(x)\,.
\ee
Hence, the number density of loops in the interval $\ell$ 
to $\ell+d\ell$ defined as $\mu\ell\,n(\ell,t)\,d\ell=\rho_{L}
(\ell,t)d\ell$ is given by 
\be
n(\ell,t)={\nu\over \ell^{5/2}t^{3/2}}\,.
\ee

At this stage we have to make an assumption as to the form of the loop
production function $f(x)$. One of the basic philosophies behind the 
scaling regime is that all the properties of the string network
are constant with respect to the horizon. Therefore, a sensible 
assumption seems to be that the loop production function is peaked 
at some constant fraction of the horizon. Mathematically, we treat this
as a delta function
\be
f(x)=c\delta(x-\alpha\zeta^{1/2})\,,
\ee
and finally one obtains 
\be
\nu=gc\alpha^{1/2}\zeta^{3/2}=g
\alpha^{1/2}
\zeta\big{(}1-\ll v^2\rr\big{)}\,.
\ee

These expressions do not take into account the effects of radiation 
on the strings loops which we discuss in the next section. 
However, this can be achieved simply using a linear decay 
of string loop energy given by $\ell=\ell_0-\kappa(t-t_0)$. Using
 $\ell_0=\alpha t_0$, one can deduce that
\be
\ell_0={\ell+\Gamma G\mu t\over 1+\Gamma G\mu/\alpha}\,.
\ee
Making the substitution $\ell\rightarrow l_0$, that is, 
$n(\ell_0,t)d\ell_0\rightarrow n(\ell,t)d\ell$, the number density of 
loops in the radiation, taking into account the decay of loop length, is 
given by 
\be
n(\ell,t)={\nu\over (\ell+\kappa t)^{5/2} t^{3/2}}
\,,\label{numdenloop}
\ee 
with  $\nu$ redefined to be
\be
\nu= \big{(}1+\kappa/\alpha\big{)}^{3/2}g
\alpha^{1/2}\zeta\big{(}1-\ll v^2\rr\big{)}\,.
\ee
All these parameters, but for the loop size $\alpha$, can be reliably estimated
from high resolution simulations of local strings\mycite{BB,AS} (for example, 
$\zeta 
\approx 13$).

\subsection{Axion mass `switch on' and domain walls}

\noindent Near the QCD phase transition the axion acquires a mass and network
evolution alters dramatically because domain walls form\mycite{Sikb}, with each
string becoming attached to a wall\mycite{VE}. `Soft' instanton 
calculations give
the initially temperature dependent mass\mycite{WilWei,Turn}
\be
m_{\rm a}(T)=(0.1\pm 0.03)\bar\ma\bigg{(}{\fa\over 10^{12}{\rm
GeV}}\bigg{)}^{-1}\bigg{(}{\Lambda_{\rm QCD}\over
T}\bigg{)}^{3.7\pm 0.1}\,,
\ee
which achieves its maximum value for $T>>\Lambda_{\rm QCD}$ at
\be
\ma= \bar\ma\bigg{(}{\fa\over 10^{12}{\rm
GeV}}\bigg{)}^{-1}\,.
\ee
This mass only becomes significant when the Compton wavelength falls inside the
horizon, that is, $m(\tilde t) \tilde t\sim 0.75$ at the time 
\be
\tilde t\sim 10.8\times 10^{-7}\Delta^2\bigg{(}{\fa\over 10^{12}{\rm
GeV}}\bigg{)}^{0.36}\bigg{(}{\bar\ma\over 6\times 10^{-6}{\rm eV}}\bigg{)}^{-2}
\bigg{(}{{\cal N_{\rm QCD}}\over 60}\bigg{)}^{0.5} \,{\rm sec}\,,
\ee 
where $\Delta$ is a constant of order unity which quantifies parameter uncertainties 
at the  QCD phase transition \footnote{Note that to 
calculate $\tilde t$, we assume an
effective number of massless degrees of freedom $\cal N$ in an epoch when its actual
value is falling. However, it should be possible to use an averaged
value 
throughout this epoch.},
\be
\Delta = 10^{\pm
0.5}\bigg{(}{\bar\ma\over 6\times 10^{-6}{\rm eV}}\bigg{)}^{0.82}
\bigg{(}{\Lambda_{\rm QCD}\over 200{\rm MeV}}\bigg{)}^{-0.65} \bigg{(}{{{\cal
N}_{\rm QCD}}\over 60}\bigg{)}^{-0.41}\,. \label{QCDuncertain}
\ee
Large field variations around the strings collapse into 
localized domain walls at $\tilde
t$.   Subsequently, these domain walls begin to dominate over 
the string dynamics
when the force from the surface tension becomes comparable to the 
tension
due to the typical string curvature $\sigma\sim\mu/t$, 
\be
t_{\rm w} \sim 1.7\times10^{-6}\Delta^2\bigg{(}{\fa\over 10^{12}{\rm
GeV}}\bigg{)}^{0.36} \bigg{(}{\bar\ma\over 6\times 10^{-6}{\rm eV}}\bigg{)}^{-2}
\bigg{(}{{\cal N}_{\rm QCD}\over 60}\bigg{)}^{0.5}\,{\rm sec}\,.
\ee
The demise of the hybrid string--wall network proceeds rapidly\mycite{VE}, as
demonstrated numerically\mycite{Sa,Sb,PRS}.  The strings frequently intersect and
intercommute with the walls, effectively `slicing up' the network into small
oscillating walls bounded by string loops.  Multiple self-intersections will reduce
these pieces in size until the strings dominate the dynamics again and decay
continues through axion emission.

\section{The nature of axion string radiation}

\noindent Axion strings oscillate relativistically and radiate their 
energy primarily into axions; this is the preferred channel because the 
string coupling to the axion is much stronger than, say, gravitational 
radiation.  The axion string is a global string with a long-range
Goldstone boson field in which most of its energy resides, as the
logarithm in (\ref{muren}) indicates. With 
$\fa \sim 10^{10}\,\GeV$ this effective renormalization is about 
$\log(t/\delta) \approx 70$, but the apparent `non-locality' of 
a global string is more
imagined than real.  To illustrate this point, consider surrounding 
a straight axion string with a narrow tube of radius $R\sim t/100$; this tube 
would contain 
95\% of the string's energy while only enclosing 0.03\% of the available
volume.  Nevertheless, the understanding of global or axion string 
dynamics has some heuristic pitfalls and, in the past, there was
considerable debate on the subject.  In order to assure the reader of 
the veracity of the conclusions that follow on this key issue,
 an appendix is provided summarizing more recent literature on the subject.

\begin{figure}
\centerline{\psfig{figure=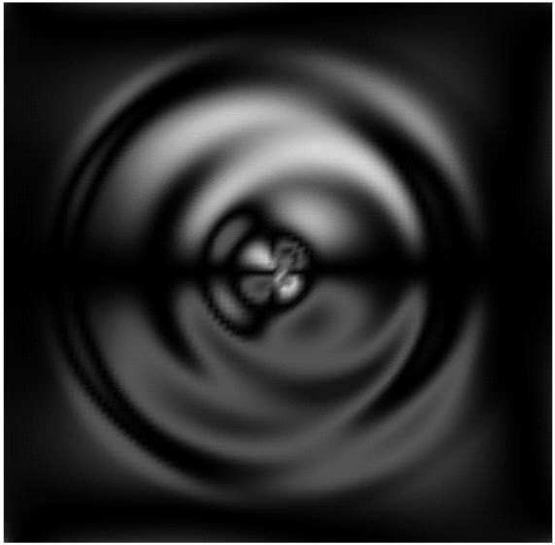,width=3.0in,height=3.0in}}
\caption{Axion radiation from an oscillating periodic string
configuration in a three-dimensional field theory simulation. In this 
perpendicular cross-section, the string oscillates horizontally.  Note
the dominance of $n=2$ quadrupole radiation.}
\end{figure}

On cosmological scales, as we have stated already, the basic fact is
that the axion string essentially behaves like a local cosmic string, 
but with stronger radiation damping.  The radiation power arising from 
an oscillating string loop is independent of it size.  This
scale-invariant power is given by
\be
P = \Gamma_{\rm a}\fa^2  \equiv \kappa\mu,\label{backscale}
\ee
where $\Gamma_{\rm a}\approx 65$ has been found from numerical
simulations and $\kappa$ essentially defines the radiation backreaction 
scale.  This radiation power loss leads to the linear decay of the loop
size with time, 
\be
\ell = \ell_{\rm i} - \kappa(t-t_{\rm i})\,.\label{loopdecay}
\ee
Any loop will disappear into axions after about 10--20 oscillations.

The radiation spectrum into which the loops decay is dominated by the 
lowest available frequencies, that is, the lowest harmonic wavelengths 
proportional to the loop size $\ell$.  If we decompose the loop
radiation  spectrum into the power in each harmonic $P_n$ then we 
expect a power law fall-off at large $n$,  that is, 
\be
P = \sum_n P_n,\qquad Pn \propto n^{-q} ~~(n>\!> 1)\,,\quad
\hbox{with}~~q\geq 4/3\,.\label{Pns}
\ee
For typical loops, the spectral index $q$ is expected to be greater than
$4/3$ because of radiative backreaction effects.  The fact that the
spectrum is dominated by the lowest harmonics can be seen in some results 
from numerical simulations of axion strings illustrated  in figs.~1--2.  For
this particular axion string configuration the quadrupole ($n=2$) is most
prominent, as shown quantitatively in fig.~2 with a power spectrum analysis.

\begin{figure}
\centerline{\psfig{figure=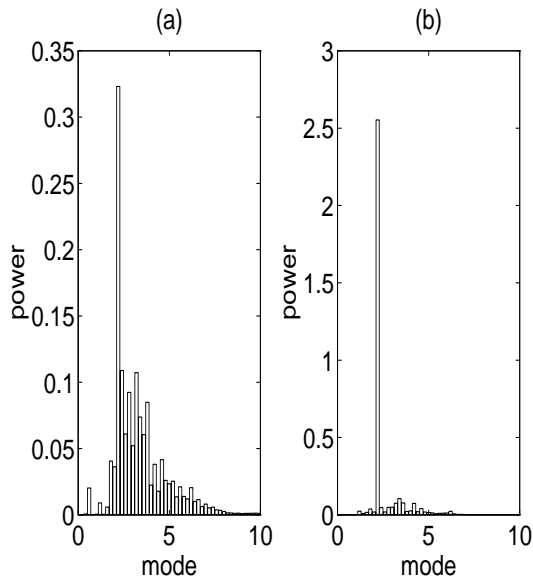,width=3.0in,height=3.0in}}
\caption{Power spectrum of axion radiation from an oscillating periodic 
string (as in fig.~2).  Here, the initial string configuration has a 
sharp `kink', but the initial high frequency radiation (a), rapidly 
gives way at later times to the dominant $n=2$ quadrupole radiation
(b).  This spectral evolution demonstrates the effect of radiative
backreaction.}
\end{figure}

The key uncertainty in cosmic axion calculations is the scale-invariant 
size $\ell = \alpha t$ at which loops are created by the string network.
The parameter $\alpha$ is determined by radiative backreaction effects
on the string network and, since the backreaction scale is set by $\kappa$
in (\ref{backscale}), most authors anticipate $\alpha \approx \kappa$.
However, smaller values are not necessarily excluded\mycite{BSa}, so here we
consider the range $0.1\lapp \alpha/\kappa \lapp 1.0$, and this key
parameter will appear in subsequent axion bounds.  The above facts are all
we need to know to calculate the cosmic axion abundance.

\section{The axion density in the standard thermal scenario}

\noindent Given the loop distribution (\ref{numdenloop}), we can calculate the energy
density of emitted axions.  The radiation spectrum from a loop of length $\ell$,
averaged over various loop  configurations, is given by 
\be
{dP_{\ell}(\omega)\over d\omega}=\fa^2\ell \, g(\ell\omega)\,,
\ee
where
the spectral function $g(x)$---a continuum limit of (\ref{Pns})---is 
normalised by
\be
\int_0^{\infty}g(x)dx=\Gamma_{\rm a}\,,\label{normalise}
\ee
and $\Gamma_{\rm a}$ is
defined in (\ref{backscale}). We shall assume that loops are at rest, because any initial velocity will
be redshifted and the net error when averaged isotropically over all loops should be
relatively small.

The energy density of massless axions emitted at time $t_1$ in an interval $dt_1$
with frequencies from $\omega_1$ to $\omega_1+d\omega_1$ is
\be
d\rho_{\rm a}(t_1)=dt_1d\omega_1\fa^2\int_{0}^{\infty}d\ell\,n(\ell,t_1)\ell\,g(\ell\omega)\,.
\ee
Assuming ${\cal N}$ constant, the spectral density can be calculated
by integrating  over $t_1<t$, taking into account the redshifting of both the
frequency, $\omega = {a(t_1)/ a(t)}\omega_1$, and the energy density, $\rho_{\rm
a}\propto a^{-4}$.  Neglecting the slow logarithmic dependence of the backreaction
scale $\kappa$,  we have 
\be
{d\rho_{\rm a}\over
d\omega}(t)={\fa^2\over t^{3/2}}\int_{t_*}^tdt_1\int_0^{\alpha
t_1}d\ell\,{\nu \ell \over (l+\kappa t_1)^{5/2}}\, g\left[\left({t/
t_1}\right)^{1/2}\omega \ell\right]\,.\label{specdensity}
\ee
Under the  substitution
$x=\ell/t_1\,,\,z=\omega x(tt_1)^{1/2}$, the range of  
integration is transformed
and (\ref{specdensity}) becomes\mycite{VS} 
\be
{d\rho_{\rm
a}\over d\omega}(t)={4\fa^2\nu\over 3\omega \kappa^{3/2}t^2} \int_{0}^{\alpha\omega
t}dz\, g(z) \left[\left(1+\left( z\over\omega \kappa t\right)\right)^{-3/2}
 - \left(1+{\alpha\over
\kappa}\right)^{-3/2}\right]\,,\label{specdensitysub}
\ee
since the contribution from the lower limit can be shown to be zero 
for the range of $\omega$ under consideration.  This implies the peak 
contribution to the 
density
comes from those axions emitted just before wall domination.

One can approximate the integrals of $g(z)$ by noting that the dominant 
contribution comes in the range $4\pi<z<4\pi n_*$, where $n_*$ is the mode beyond
which the radiation spectrum of loops can be truncated due to backreaction. Assuming
$4\pi n_*\less\omega \kappa t$ and $\alpha\lapp\kappa$, one 
can use the normalisation
condition  (\ref{normalise}), to deduce that the integral (\ref{specdensitysub}) becomes 
\be
{d\rho_{\rm a}\over d\omega}(t)\approx {4\Gamma_{\rm
a}\fa^2\nu \over 3\omega
\kappa^{3/2}t^2}\left[1-\left(1+{\alpha\over\kappa}\right)^{-3/2}\right]\,.
\ee
This estimate is only formally accurate for $\alpha\lapp\kappa$, but it should also yield
useful information for $\alpha\gapp\kappa$. From this expression we can obtain
the spectral number density of axions 
\be
{dn_{\rm a}\over d\omega}={1\over\omega}{d\rho_{\rm
a}\over d\omega}\,.
\ee
Integrating  and comparing with the entropy density of the universe,
$s=2\pi^2{\cal N}T^3/45$, the ratio of the axion number density to the entropy at
$t_{\rm w}$ can be calculated as  
\bea
{n_{\rm a}\over s}\approx &&
6.7\times10^6 \left( 1+{\kappa\over\alpha}\right)^{3/2}\left[1
 - \left(1+{\alpha\over \kappa}\right)^{-3/2}\right]\\ && \Delta\bigg{(}{\bar\ma\over 6\times 10^{-6}{\rm eV}}\bigg{)}^{-1}\bigg{(}{\fa\over 10^{12}{\rm
GeV}}\bigg{)}^{2.18}\,,
\eea
using typical parameter values $\Gamma_{\rm a}\approx 65$
and
$\nu\approx 6\alpha^{1/2}(1+\kappa/\alpha)^{3/2}$.
Assuming number conservation after $t_{\rm w}$ and using the entropy density
$s_{0}=2809(T_{0}/2.7{\rm K})^3{\rm cm}^{-3}$ and critical density $\rho_{\rm
crit}=1.88h^2\times10^{-29}{\rm g cm}^{-3}$ at the present day, one can deduce  that
the axion loop contribution is 
\be
\Omega_{\rm a,\ell}\approx 10.7
\bigg{[}\bigg{(}1+{\alpha\over\kappa}\bigg{)}^{3/2}-1\bigg{]}
h^{-2}\Delta\bigg{(}{T_{0}\over 2.7{\rm K}}\bigg{)}^{3}\bigg{(}{\fa\over 10^{12}{\rm
GeV}}\bigg{)}^{1.18}\,.\label{finalcont}
\ee
It should
be noted that the dependence of (\ref{finalcont}) on the ratio $\alpha/\kappa$ comes
about because the lifetime of a loop produced at $t_{\rm i}$ is
$(\alpha/\kappa)t_{\rm i}$.

The contribution from long strings can also be roughly estimated assuming that the
radiation from the long string network does not affect the scaling balance condition.
The basis for this calculation is the radiation power per unit length 
\be
{dP\over d\ell} \approx {\pi^3\fa^2\over 16\gamma t}\,,
\ee
with the long string backreaction scale given by  $\gamma \sim
(\pi^2/8)[\ln (t/\delta)]^{-1}$. Assuming the radiative dominance of this 
smallest
scale $\gamma t$ (as noted elsewhere in simulations\mycite{AShe2}), 
one can calculate the spectral
density of axions from long strings  
\be
{d\rho_{\rm a}\over
d\omega}\approx {\pi^3\fa^2\zeta\over 8\gamma\omega t^2}\,.
\ee
Using similar methods to those used for loops we obtain
\be
\Omega_{\rm
a,\infty}\approx 1.2 h^{-2}\Delta\bigg{(}{T_{0}\over 2.7{\rm
K}}\bigg{)}^{3}\bigg{(}{\fa\over 10^{12}{\rm GeV}}\bigg{)}^{1.18}\,,
\label{finallongstring}
\ee
which is found to be independent of the actual backreaction
scale $\gamma$.   The considerable uncertainty of (\ref{finallongstring}) must be
emphasised given its sensitivity to the amplitude of small-scale structure and the
overall long string radiation spectrum.   A comparison of the two contributions
(\ref{finalcont}) and (\ref{finallongstring}) yields, 
\be
{\Omega_{\rm
a,\ell}\over \Omega_{\rm a,\infty}}\approx
10.9\bigg{[}\bigg{(}1+{\alpha\over\kappa}\bigg{)}^{3/2}-1\bigg{]}\,,
\ee
independent of $\Delta$ and $h$. For the expected parameter range, that is
$0.1<\alpha/\kappa <1.0$, the loop contribution is considerable larger.
We discuss the axion bound quantitatively in the conclusion.

An order-of-magnitude estimate of the demise of the string/domain wall
network\mycite{Lyth2} indicates that there is an additional contribution
\be
\Omega_{\rm a,dw}\sim{\cal O}(1)h^{-2}\Delta\bigg{(}{T_0\over 2.7{\rm
K}}\bigg{)}^3\bigg{(}{\fa\over 10^{12}{\rm GeV}}\bigg{)}^{1.18}\,.
\ee
This `domain
wall' contribution is ultimately due to loops which are created at the time $\sim
t_{\rm w}$. Although the resulting loop density will be similar to (\ref{numdenloop}),
there is not the same accumulation from early times, so it is likely to be
subdominant relative to (\ref{finalcont}). Both the long string and domain wall
contributions will serve to strengthen the loop bound (\ref{finalcont}) on the axion.

\section{Alternative scenarios}

\subsection{Inflation}

\noindent The discussion of the axion density presented  so far presumes that a
global string network forms after any epoch of inflation, that is, $T_{\rm
reh}>\fa$ in the standard thermal scenario (i) of section 1.3.
In the alternative scenario when $T_{\rm reh}<\fa$, the axion density
and any strings formed before inflation are exponentially
suppressed by the rapid expansion\footnote{It is possible to form strings during
inflation\mycite{Lyth2}, however, the significance of this possibility is far 
from
clear at the present, since quantitative predictions of the initial 
conditions are difficult to make. 
If just a small number of long strings survive 
in the initial distribution, then it may be that the scaling regime 
can be achieved, 
albeit slowly. Since the most important axions are those emitted 
just before the QCD phase transition, 
the network will have plenty time
to reach the appropriate scaling regime.}. In
this case, the only contribution to the axion density comes from the initial
misalignment mechanism, that is, $\Omega_{\rm a}$ as given by (\ref{homcont}). However,
$\theta_{\rm i}$ is homogeneous on scales larger then the current horizon  and
there is no a priori reason to suppose that $\theta_{\rm i}$ should take any
particular value. Essentially, with the freedom to choose any value of $\theta_i$,
there is no constraint on $\fa$. Some authors\mycite{Lindeb} seem to favour larger
values of $\theta_{\rm i}(\sim \pi/2)$ as being more `natural', since they avoid
apparent fine tuning and anthropic arguments about our region of the 
universe. For example, if one were to 
chose to conservatively limit $\theta_{\rm i}$ to lie in the range
$0.1\lapp\theta_{\rm i}\lapp \pi /2$, then the constraint on the axion becomes 
$\fa\,\lapp\,
10^{11}\hbox{--}10^{14}{\rm GeV}\,,~(\ma\,\gapp\, 0.05\hbox{--}50\mu{\rm eV})$, 
which is---it has to 
be admitted---a rather indefinite prediction (before other particle 
physics and cosmological uncertainties 
are included).   Further details about this scenario can be found in general
reviews\mycite{axreviews}.

Various attempts have been made to avoid the anthropic arguments associated 
with this `quiescent' inflationary axion by
appealing to particle physics-motivated models in which 
$\theta_{\rm
i}$ is set by the conditions at the end of inflation, with one possibility 
being hybrid inflation\mycite{Linde}. In this case, the axion field is coupled to 
the
inflaton causing the formation of topological defects at the end of inflation,
but this returns us to the standard thermal scenario described previously. 
Such  models may occur naturally in supersymmetric axion
models\mycite{LythStew}.  

\subsection{Entropy production}

Entropy production by the out-of-equilibrium decay  of massive particles between
the QCD phase transition and nucleosynthesis, can weaken all the bounds on $\fa$.
If the entropy is increased by some factor $\beta$, that is $s\rightarrow \beta s$,
then the axion density is decreased by a factor $\beta^{-1}$, that is
$\Omega_{\rm a}\rightarrow \beta^{-1}\Omega_{\rm a}$. For example, it has been
noted that that the decay of the saxino---the spin zero partner of the 
axion---can lead to a dilution by\mycite{Lyth1}  
\be
\beta < 5\times
10^{3}\bigg{(}{m_{\rm sax}\over 1{\rm TeV}}\bigg{)}\,,
\ee which can be
up to a factor of 1000.  Such entropy production appears to be a
substantial and somewhat inelegant extension of the axion model which 
combines an extra degree of fine-tuning. However, it is another
mechanism by which to produce an axion detectable in the current search 
range $1\hbox{--}10\,\microeV$.

\section{Conclusions}

\begin{figure}
\centerline{\psfig{figure=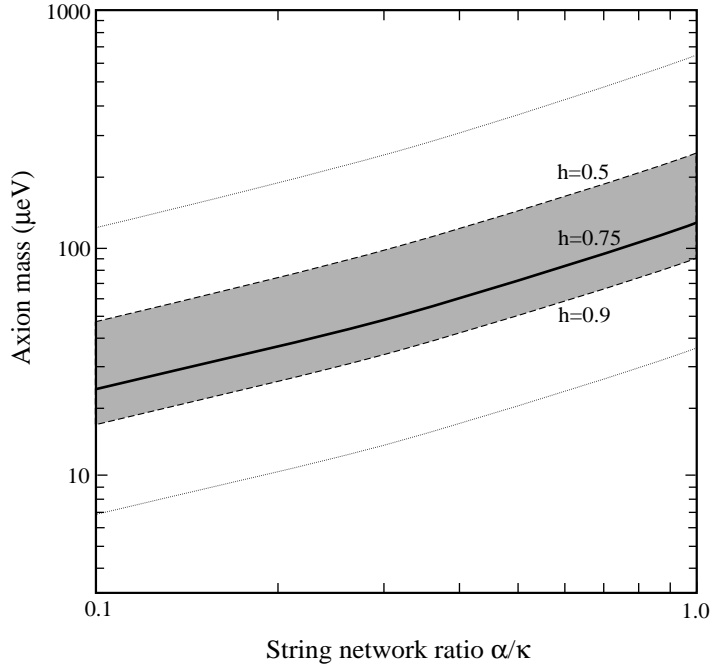,width=4.0in,height=3.5in}}
\caption{Parameter uncertainties in estimates of the dark matter axion
mass.  The key string parameter $\alpha/\kappa$ (the 
loop-size/backreaction-scale ratio) is plotted from the usual assumed value 
$\alpha/\kappa\approx 1$
to a possible lower limit $\alpha/\kappa = 0.1$.  The bold line is the 
axion mass for a Hubble parameter $h=0.75$ and the shaded region allows
for errors in the range $0.5\lapp h\lapp 0.9$.  The dotted line
compounds this with particle physics uncertainties.
\label{fig:new_ma}}
\end{figure}

\noindent The cosmological axion density has been
calculated in the standard thermal scenario by considering the dominant
contribution from axion strings.  Unlike the alternative scenarios
described above, there is, in principle, a well-defined calculational
method to precisely predict the mass $\ma$ of a dark matter axion.
For the currently favoured values of the Hubble parameter ($H_0 \approx 
75\rm  \,km\,s^{-1}Mpc^{-1}$), we use string 
network simulation parameters\mycite{BB,AS}
in (\ref{finalcont}) to find
\bea
&\fa ~\lapp~  4.7 \times 10^{10}
\,{\rm GeV}\,\qquad\ma ~\gapp~ 130\,\mu{\rm eV}\,,\qquad h=0.75\,,\ \\
&~[\fa ~\lapp~  2.3 \times 10^{10}
\,{\rm GeV}\,\qquad\ma ~\gapp~ 250\,\mu{\rm eV}\,,\qquad h=0.5]\,,\ \\
&[\fa ~\lapp~  6.4
\times 10^{10} \,{\rm GeV}\,\qquad\ma ~\gapp~~ 78\,\mu{\rm eV}\,,\qquad
h=0.9].
\eea
The key uncertainty in this string calculation is the parameter ratio
$\alpha/\kappa$ (that is, the loop-size to backreaction scale ratio).
Here, like most authors 
we have assumed $\alpha/\kappa\approx1.0$, however, this remains to be 
firmly established quantitatively.  Conceivably, a smaller ratio as low as 
 $\alpha/\kappa\approx0.1$ is possible and could weaken the bound to 
$\ma \gapp 24\,\microeV$ and $\fa \lapp 2.5\times10^{11}\GeV$ for
$h=0.75$.  We exhibit the effect of $\alpha/\kappa$ on the axion mass in
fig.~3.  Reducing this uncertainty remains a key research goal, though a
technically difficult one.  Also included in fig.~3 are the theoretical
particle physics uncertainties, which are about a further factor of 2.5.

Even the weakest axion string bound is considerably stronger than
the original constraint from the `quiescent' production of zero momentum
axions for which $\fa \,\lapp\, 10^{12}\GeV$.  This bound 
narrows the axion window 
but there still remains a considerable range, $\fa \sim 10^9\hbox{--}10^{11}$GeV, 
lying above the astrophysical constraint.  
These estimates motivate the search for a slightly heavier axion 
outside the detectable range of the present axion dark matter experiment.  
Pinpointing more precisely the 
predicted dark matter 
axion mass near $\ma \sim 100\,\microeV$ remains a theoretical
priority.

\section*{Acknowledgments}

EPS is particularly grateful to Professor Klapdor-Kleingrothaus and 
to the other Heidelberg workshop organisers for their invitation and  
hospitality.
We would both like to thank Alex Vilenkin, Rob Caldwell, Michael Turner, 
Georg Raffelt and David Lyth for helpful discussions during this work. 
RAB is supported by PPARC postdoctoral fellowship grant GR/K94799. This 
work has also been supported by PPARC rolling grant GR/K29272.

\section{Appendix: Axion string radiation}

\subsection{The Goldstone model}

\noindent The essential features of global strings are exhibited in the simple
U(1) Goldstone model with action given by 
\bea
&S = \int d^4x \bigg{[}\partial_{\mu}\bar\Phi\partial^{\mu}\Phi
-{1\over 4}\lambda(\bar\Phi\Phi - \fa^2 )^2 \bigg{]} \\ = &\int d^4x \bigg{[}(\partial_{\mu}\phi)^2 + \phi^2 (\partial_\mu\vartheta)^2
-{1\over 4}\lambda(\phi^2 - \fa^2 )^2 \bigg{]}\,,\label{actionone}
\eea
where $\Phi$ is a complex scalar field which has been split as
$\Phi = \phi e^{i\vartheta}$ into a massive (real) Higgs component $\phi$ and a
massless (real periodic) Goldstone boson $\vartheta$.  

For a straight global string in flat space,
lying along the $z$-axis, the appropriate ansatz is 
\be
\Phi(r,
\theta) = \phi(r) e^{in\theta}\,,
\ee
where  $\theta$ is the azimuthal
angle and $n$ is the winding number, and one takes the usual boundary conditions,
$\phi(0) =0$ and $\phi\rightarrow\fa $ as $ r\rightarrow\infty$. Despite these
conditions, the energy per unit length $\mu$ is logarithmically divergent, 
\be
\mu(\Delta) ~\approx~ \mu_0 + 
\int_\delta^\Delta \left [ {1\over r}{\partial \Phi\over \partial
\theta}\right]^2 2\pi r\, dr ~\approx~\mu_0 + 2\pi\fa^2 \log \left (\Delta\over
\delta\right)\,,\label{stringxsect}
\ee
where $\delta \sim (\sqrt \lambda \,\fa)^{-1}$ is
the string core width and $\mu_0\sim \fa^2$ is the core energy associated with
the massive field $\phi$ (that is, within $r \lapp \delta$).  The length-scale
$\Delta$ is the renormalisation scale provided in general by the curvature radius
of the string or the average inter-string separation.

It is clear from this that the string is not localised.
However, in the following discussion it will be shown that the dynamics of 
global strings are similar to those of other types of strings, except for this
renormalised string tension.
Given $\fa \sim 10^{10}-10^{12}$GeV, the logarithm in (\ref{stringxsect}) 
implies that there is much more 
energy in the  Goldstone field than in the string core $\mu_0$.  Thus  
the dynamics 
of the string are dictated by the Goldstone self-field, a 
fact which has made the understanding of global strings intuitively hazardous.

\subsection{The Kalb-Ramond Action}

\noindent The analytic treatment of global
string dynamics is hampered by the topological coupling of the string to the
Goldstone boson radiation field.  However, we can exploit the  well-known duality
between a massless scalar field and a two-index antisymmetric tensor field
$B_{\mu\nu}$ to replace the Goldstone boson $\vartheta$ in (\ref{actionone}) via
the relation 
\be
\phi^2\partial_{\mu}\vartheta = {1 \over 2}f_a\epsilon_{\mu\nu\lambda\rho}\partial
^{\nu}B^{\lambda\rho}\,. \label{duality}
\ee
The canonical transformation generated by (\ref{duality}) requires the addition of
a total derivative to the action. In this case the total derivative which generates
the transformation is given by\mycite{DSb,W,VV}
\be
\delta S=\int\nolimits
d^4x\,\epsilon^{\mu\nu\lambda\rho}\partial_{\nu}B_{\nu\lambda}\partial_{\rho}\vartheta =\int d^4x\,\epsilon^{\mu\nu\lambda\rho}\bigg{[}\partial_{\mu}(B_
{\nu\lambda}\partial_{\rho}\vartheta)-B_{\nu\lambda}\partial_{\mu}\partial_{\rho}\vartheta
\bigg{]}\,\label{transaction}
\ee
where $\epsilon^{\mu\rho\nu\lambda}$ is the totally antisymmetric tensor in 
four dimensions.
Under this transformation the Goldstone action (\ref{actionone}) becomes 
\be
S = 
\int d^4x\bigg{[}(\partial_{\mu}\phi)^2
+{f_a^2 \over 6\phi^2}H^2 - {1 \over 4}\lambda(\phi^2 - f_a^2)^2\bigg{]} 
 \,,\label{prekraction}
\ee
where the field strength tensor is $H^{\mu\nu\lambda} = \partial^{\mu}B^{\nu\lambda} +
\partial^{\lambda}B^{\mu\nu} +  \partial^{\nu}B^{\lambda\mu}$. One should note 
that the sign of the $H^2$ term is the opposite of that one would deduce by
substituting  
the duality relation (\ref{duality}) directly into the Goldstone action (\ref{actionone}). 
One can deduce the correct sign, as shown in (\ref{prekraction}), by treating 
the antisymmetric tensor as a Lagrange multiplier.

In the case of a spontaneously broken symmetry, the term added to the action (\ref{transaction}) is no longer a
total derivative due to the presence of a vortex at $\phi=0$. One finds that the
commutator $\epsilon^{\mu\rho\nu\lambda}\partial_{\mu}\partial_{\rho}\vartheta$ is non
zero and one can define  
\be
\epsilon^{\mu\rho\nu\lambda}\partial_{\mu}\partial_{\rho}\vartheta=4\pi
J^{\nu\lambda} \,,\label{current}
\ee
where $J^{\nu\lambda}$ is an effective current
density given by 
\be
J^{\mu\nu} = {\fa\over
2}\int\delta^{(4)}\big{(}x-X(\sigma,\tau)\big{)}\,
d\sigma^{\mu\nu}\,.
\ee
The area element $d\sigma^{\mu\nu}$ is given in
terms of the  worldsheet $X(\sigma,\tau)$ swept out by the zeroes of the Higgs
field ($\phi=0$),  
\be
d\sigma^{\mu\nu} = \epsilon^{ab}
\partial_aX^{\mu}\partial_bX^{\nu}d\sigma d\tau\,.
\ee
Rearranging the
terms and evaluating the delta function in (\ref{current}) allows the action to be
written as,
\be
S = \int\nolimits
d^4x\,\bigg{[}(\partial_{\mu}\phi)^2 +{f_a^2 \over
6\phi^2}H^2 - {1 \over 4}\lambda(\phi^2 - f_a^2)^2   \bigg{]} - 2\pi
f_a\int\nolimits B_{\mu\nu}d\sigma^{\mu\nu}\,.
\ee

As for strings for local strings,
in the neighbourhood
of the string, one may use a local coordinate system to define,
\be
x^{\mu}(\sigma,\tau,\rho^1,\rho^2)=X^{\mu}(\sigma,\tau)+m^{\mu}_a\rho^a\,,
\ee
where
$X^{\mu}(\sigma,\tau)$ are the coordinates on the string worldsheet and
$m_a^{\mu}$ are two orthonormal vectors perpendicular to the worldsheet and 
$\rho^{a}$ are related coordinates. If one
integrates radially over the massive degrees of freedom by defining the bare
string tension  to be  
\be
\mu_0 = -\int
d^2\rho\bigg{[}(\partial_{\mu}\phi)^2+{1\over 6}
\bigg{(}{\fa^2\over\phi^2}-1\bigg{)}H^2-{1\over
4}\lambda(\phi^2-\fa^2)^2\bigg{]}\,,
\ee
one can deduce the
Kalb-Ramond action\mycite{KR}, 
\be
S  = -\mu_0\int\nolimits \sqrt{-\gamma} \, d\sigma d\tau +
{1 \over 6}\int\nolimits d^4x\, H^2 - 2\pi f_a\int\nolimits
B_{\mu\nu}d\sigma^{\mu\nu}\,.\label{kraction}
\ee 
where $\gamma_{ab}=g_{\mu\nu}\partial_a X^{\mu}\partial_b X^{\nu}$ is the
induced metric on the worldsheet and $\gamma={\rm det}(\gamma_{ab})$. The first term
is the well known Nambu action of local strings, the
second is the antisymmetric tensor field strength for both external fields and the
self field of the string and the last is a contact interaction between the
$B^{\mu\nu}$ field and the string worldsheet. This coupling between the string and $B_{\mu\nu}$ is analogous to the
electromagnetic coupling of a charged particle, and is amenable to the same
calculational techniques. This is the basis for the subsequent analytic work reviewed here.

\subsection{Equations of motion}

\noindent Varying the Kalb-Ramond action (\ref{kraction}) with respect to the worldsheet
coordinates and the antisymmetric tensor,
gives the equations of motion for the string and  the antisymmetric tensor
field equation,  
\bea
&\mu_0\partial_a\big{(}\sqrt{-\gamma}\gamma^{ab}\partial_bX^{\mu}\big{)}
=  {\cal F}^{\mu} = 2\pi\fa H^{\mu\alpha\beta}\epsilon^{ab}\partial_aX_{\alpha}
\partial_bX_{\beta}\,,\\
&\partial_{\mu}H^{\mu\alpha\beta}=-4\pi J^{\alpha\beta}=-2\pi\fa\int d\sigma
d\tau\,\delta^{4}\big{(}
x-X(\sigma,\tau)\big{)}\,\epsilon^{ab}\partial_aX^{\alpha}\partial_bX^{\beta}
\,.
\eea
One can use the conformal string gauge 
${\dot X}^2+X^{\prime 2}=0$ and ${\dot X}\cdot X^{\prime}=0$,
and also the
Lorentz gauge of the antisymmetric tensor field, that is, $\partial_{\mu}
B^{\mu\nu}=0$, to deduce that
\bea
&\mu_0(\ddot X^{\mu} - X^{\prime\prime\mu}) = 
{\cal F}^{\mu}= 2\pi
f_aH^{\mu\alpha\beta} (\dot X_{\alpha} X^{\prime}_{\beta} -
X^{\prime}_{\alpha}\dot X_{\beta})\,,\\ &\square B_{\alpha\beta} = 2\pi f_a
\int\nolimits d\sigma d\tau(\dot X_{\alpha} X^{\prime}_{\beta} -
X^{\prime}_{\alpha}\dot X_{\beta}) \delta^{(4)}\big{(}x-X(\sigma,\tau)
\big{)}\,.
\eea 
As for the point electron one can split up the force ${\cal F}^{\mu}$ into two parts; one due to the self-field and the finite radiation backreaction force. The equations of motion are therefore
\be 
\mu_0(\ddot X^{\mu} - X^{\prime\prime\mu}) = {\cal F}^{\mu}_{\rm s}+{\cal F}^{\mu}_{\rm r}\,,
\ee
where the self-force is 
\be
{\cal F}^{\mu}_{\rm s}= -2\pi\fa^2\log(\Delta/\delta)\left[\ddot X^{\mu} -
X^{\mu\prime\prime}\right]
\ee
and a first-order approximation to the finite radiation backreaction force is 
given by\mycite{BSc,BSd}
\be 
{\cal F}^{\mu}_{\rm r}= {4\pi\fa^2\Delta\over}\left[x^{\tdot\mu}-\left(
{\dot X\cdot x^{\tdot}\over \dot X^2}\right)\dot X^{\mu} + \left({
X^{\prime}\cdot X^{\tdot}\over \dot X^2}\right)X^{\mu\prime}\right]\,.
\ee
where $\Delta$ is the renormalisation scale introduced in
(\ref{stringxsect}).
The self field is
logarithmically divergent, but it can be absorbed such that the equations of
motion in the conformal gauge become\mycite{LR,DQ}
\be 
\mu(\Delta)\big{(}\ddot X^{\mu} - X^{\prime\prime\mu}) =
{\cal F}^{\mu}_{\rm r}\,.\label{renormeom}
\ee
The renormalised equations of motion (\ref{renormeom}) for the
string can be approximated by the Nambu equations of motion, assuming the effects
of radiation backreaction to be small, that is $F^{\mu}_{\rm r}\approx 0$. 

\begin{figure}
\centerline{\psfig{figure=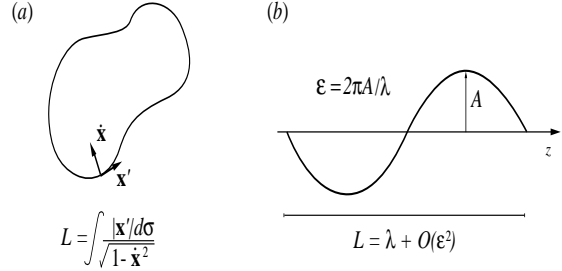,width=3.0in,height=1.5in}}
\caption{Loop and periodic long string trajectories which have been the 
subject of extensive analytic and numerical studies.
\label{fig:loop_long}}
\end{figure}

These  equations have closed loop and periodic long (or infinite) string solutions. The loop solutions are parametrized by their invariant length of the loop $L$, which is closely related to the characteristic frequency  $\Omega=2\pi/L$, whereas the
long periodic solutions are parametrized 
by their wavelength $L$ and the ratio
of amplitude to wavelength or the relative amplitude ${\cal E}=2\pi A/L$,
where $A$ is their amplitude. Fig.~\ref{fig:loop_long} shows a schematic of the two types of
solution we have considered in detail.

\subsection{Appendix conclusion: The nature of string radiation}

\noindent The nature of 
radiation from oscillating axion (and local) strings has been extensively 
studied\mycite{VVb,VV,axrad,BSa}. The methods developed
allow one, in principle, to calculate the radiation power from any 
arbitrary periodic
loop or long string configurations, yielding the magnitude of the total power
as well as the spectrum 
of the radiation. For arbitrary long string configurations we have also
derived a linearized expression which allows one to calculate  the leading order
contribution to the radiation power\mycite{BSc,BSd}. By thorough examination of the expressions
for the radiation power and applying them to various loop and long string
configurations, we have been able to make the following deductions:
\smallskip
\noindent (i) The radiation power for loops is independent of the loop
size\mycite{VV,VVb} with the power given by
$P=\Gamma_a\fa^2=\kappa\mu$
which leads to the linear decay in the loop size $\ell=\ell_0-\kappa(t-t_0)$
as in (\ref{loopdecay}).
\smallskip
\noindent (ii) Loops radiate principally in their 
fundamental mode with a radiation
spectrum such that $P_n\propto n^{-q}$ for  large
$n$ where $q>4/3$.
\smallskip
\noindent (iii) The radiation from long strings is generically in the 
fundamental mode. However, for exactly periodic solutions the 
fundamental mode is suppressed and the radiation is quadrupole (as 
in figs.~1--2).
\smallskip
\noindent (iv) The radiation power from long string configurations 
is proportional to  ${\cal E}^2$, 
which leads to the exponential decay
of their amplitude ${\cal E}$. 
Again, for exactly periodic configurations cancellation 
gives radiation proportional to ${\cal E}^4$, leading to a 
power-law type decay. 
\smallskip
\noindent (v) Some string configurations, like kink solutions, 
which theoretically have a pathological divergence in their
radiation  power ($P_n\propto n^{-1}$),
rapidly revert to a smooth spectrum dominated by the lowest harmonics 
because of radiation backreaction effects (see fig.~2).

\smallskip
\noindent These predictions broadly uphold the assumptions behind the 
original
work on global string radiation\mycite{Dava,Davb,DSa} 
but they are completely contrary to the predictions of a `harder' $P_n \propto n^{-1}$ 
spectrum\mycite{Sikb,HarS,HagS}. 
We conclude that axion
radiation from axion
strings, when extrapolated to cosmological scales, will be 
very similar to gravitational radiation from Nambu strings.


\def\jnl#1#2#3#4#5#6{\hang{#1, {\it #4\/} {\bf #5}, #6 (#2).}}


\def\jnlerr#1#2#3#4#5#6#7#8{\hang{#1, {\it #4\/} {\bf #5}, #6 (#2).
{Erratum:} {\it #4\/} {\bf #7}, #8.}}


\def\jnltwo#1#2#3#4#5#6#7#8#9{\hang{#1, {\it #4\/} {\bf #5}, #6 (#2);
{\it #7\/} {\bf #8}, #9.}}

\def\prep#1#2#3#4{\hang{#1 [#2],  #4.}}

\def\myprep#1#2#3#4{\hang{#1 [#2], '#3', #4.}}

\def\proc#1#2#3#4#5#6{\hang{#1 [#2], `#3', in {\it #4\/}, #5, eds.\ (#6).}
}
\def\procu#1#2#3#4#5#6{\hang{#1 [#2], in {\it #4\/}, #5, ed.\ (#6).}
}

\def\book#1#2#3#4{\hang{#1 [#2], {\it #3\/} (#4).}
									}

\def\genref#1#2#3{\hang{#1 [#2], #3}
									}


\def\prl{Phys.\ Rev.\ Lett.}
\def\pr{Phys.\ Rev.}
\def\pl{Phys.\ Lett.}
\def\np{Nucl.\ Phys.}
\def\prp{Phys.\ Rep.}
\def\rmp{Rev.\ Mod.\ Phys.}
\def\cmp{Comm.\ Math.\ Phys.}
\def\mpl{Mod.\ Phys.\ Lett.}
\def\apj{Ap.\ J.}
\def\apjl{Ap.\ J.\ Lett.}
\def\aap{Astron.\ Ap.}
\def\cqg{Class.\ Quant.\ Grav.} 
\def\grg{Gen.\ Rel.\ Grav.}
\def\mn{M.$\,$N.$\,$R.$\,$A.$\,$S.}
\def\ptp{Prog.\ Theor.\ Phys.}
\def\jetp{Sov.\ Phys.\ JETP}
\def\jetpl{JETP Lett.}
\def\jmp{J.\ Math.\ Phys.}
\def\cupress{Cambridge University Press}
\def\pup{Princeton University Press}
\def\wss{World Scientific, Singapore}

\end{document}